\documentclass{aastex631}
\usepackage{url}
\usepackage{bm}
\usepackage{tabularx}
\usepackage[none]{hyphenat}

\newcommand\clearrow{\global\let\rowmac\relax}
\usepackage{float}
\usepackage{graphicx}
\usepackage{verbatim} 
\usepackage{xcolor,soul}
\usepackage{amsmath}
\definecolor{pastelyellow}{rgb}{0.99, 0.99, 0.59}

\begin{document}

\title{3D Radiative MHD Modeling of Particle Beam Heating of the Solar Atmosphere}

\author{Samuel Granovsky}
\affiliation{Department of Physics, New Jersey Institute of Technology, Newark, NJ 07102, USA}

\author{Alexander G. Kosovichev}
\affiliation{Department of Physics, New Jersey Institute of Technology, Newark, NJ 07102, USA}
\affiliation{NASA Ames Research Center, Moffett Field, Mountain View, CA 94035, USA}

\author{Irina N. Kitiashvili}
\affiliation{NASA Ames Research Center, Moffett Field, Mountain View, CA 94035, USA}

\author{Alan A. Wray}
\affiliation{NASA Ames Research Center, Moffett Field, Mountain View, CA 94035, USA}

\begin{abstract}
While solar flares are primarily associated with enhanced ultraviolet and X-ray emission, a subset of flares exhibit significant continuum brightening in visible light and are classified as white-light flares (WLFs). Despite extensive observational and modeling efforts, the physical mechanisms responsible for the compact, short-lived photospheric brightenings in WLF kernels observed during the impulsive phase of solar flares remain uncertain. Thick-target electron-beam models typically deposit energy in the upper chromosphere, and their ability to reproduce the magnitude and spatial localization of photospheric continuum enhancements observed in white-light flare kernels remains an open question. To investigate the role of self-consistent atmospheric structuring and multidimensional hydrodynamic transport in flare energy deposition, we perform three-dimensional radiative MHD simulations of electron-beam heating using the StellarBox code for a beam energy flux density of $10^{12}$ erg\,s$^{-1}$\,cm$^{-2}$ and low-energy cutoffs of 10--25\,keV. We then compute Fe\,I 6173\,\AA~Stokes profiles using the RH 1.5D radiative transfer code for direct comparison with Helioseismic and Magnetic Imager (HMI) observations. The simulations produce strong upper-chromospheric heating, multiple shock fronts, and continuum enhancements up to a factor of 2.5 relative to pre-flare levels, comparable to continuum enhancements observed during strong X-class white-light flares. Comparison with one-dimensional RADYN simulations highlights the influence of fine-scale structuring on flare dynamics and continuum emission that arises in three-dimensional geometry.
 
\end{abstract}


\section{Introduction}
Solar flares release large amounts of energy, observed across the electromagnetic spectrum. While flare radiation is most pronounced in X-ray and ultraviolet bands, a subset of flares produce distinct increases in visible continuum intensity and are known as white-light flares (WLFs). The compact, short-lived bright kernels of WLFs are understood to trace energy deposition in the low chromosphere and upper photosphere, where the bulk of the energy from typical flare electron distributions is deposited in the chromosphere. Observations of the Helioseismic and Magnetic Imager (HMI) onboard the Solar Dynamics Observatory (SDO) \citep{Scherrer2012} that show Fe\,I 6173\,\AA~line core emission \citep{Kosovichev2023} together with helioseismic signatures (sunquakes) provide direct evidence for impulsive impacts at photospheric depths and motivate detailed modeling that couples particle transport, radiative transfer, and magneto-convection.

In the standard thick-target framework \citep{1971SoPh...18..489B, 1972SoPh...24..414H, 1972SvA....16..273S}, non-thermal electrons accelerated in the corona precipitate along magnetic field lines and dissipate their energy through Coulomb collisions in the upper chromosphere. Early hydrodynamic studies \citep{1975SvA....18..590K, Livshits1981} demonstrated that this process drives a strong upward evaporation front and a downward-moving chromospheric condensation, driven by pressure gradients associated with impulsive energy deposition, but leaves the deeper photosphere largely unaffected. The predicted heating is insufficient to account for the strong continuum brightenings, Fe\,I 6173\,\AA~line core emission, and photospheric impulses inferred from helioseismic observations of sunquakes. Radiative back-warming of deeper layers \citep{1989SoPh..124..303M} may enhance the continuum emission. Yet, even with this effect, conventional electron-beam models typically underestimate both the amplitude and duration of observed white-light kernels.

One-dimensional radiative-hydrodynamic (RHD) modeling using the RADYN code has provided essential insight into flare energy deposition and chromospheric response \citep{Allred2005, 2015ApJ...809..104A, Carlsson2023}. In the F-CHROMA grid of RADYN flare models \citep{Carlsson2023}, a publicly available suite of 1D radiative hydrodynamic simulations spanning a wide range of electron-beam energy flux densities, spectral power-law indices, and low-energy cutoffs ($E_C$), electron beams with energy flux densities of $10^{11}$ to $10^{12}$ erg\,s$^{-1}$\,cm$^{-2}$, power-law indices $\delta$ of 3 to 5, and $E_C$ of 15--25\,keV produce upper-chromospheric heating and modest continuum enhancements of a few percent. In these F-CHROMA models, the energy deposition remains confined above about 1\,Mm, and the transition region is represented by a relatively abrupt rise from about $10^4$\,K to $10^6$\,K near z $\approx$ 2\,Mm. While these 1D simulations reproduce many chromospheric observables, they do not generally reproduce the strong photospheric heating and Fe\,I 6173\,\AA~emission observed by HMI during intense WLFs \citep{Sadykov2020, Monson2021}. Moreover, their smooth, semi-empirical initial atmospheres lack the fine magnetic and thermal structuring evident in flare footpoints, limiting their ability to model spatially localized energy deposition.

Recent work has begun to incorporate non-thermal electron beams directly into three-dimensional radiative magnetohydrodynamic simulations. \cite{2020A&A...643A..27F} introduced a framework in which electron beams are accelerated in reconnecting regions and propagated along magnetic field lines in a quiet-Sun atmosphere, demonstrating that beam heating can dominate conductive transport below the transition region and produce localized chromospheric heating strands. Subsequent extensions of this model \citep{2024A&A...683A.195F} incorporated magnetic gradient forces, return currents, and additional transport effects, showing that magnetic convergence and mirroring can significantly modify the spatial distribution and depth of beam energy deposition. These studies emphasize self-consistent particle transport in structured coronal magnetic geometries, but focus primarily on the coronal and upper-chromospheric energy balance rather than the detailed hydrodynamic and radiative response of the lower atmosphere.

Parallel efforts have expanded the RHD framework to include proton and mixed electron-proton beams as alternative energy carriers. Proton beams penetrate to greater column depths than electrons and can produce both white-light emission and helioseismic signatures. In particular, \citep{Sadykov2024RADYN} presented a comprehensive suite of RADYN + Fokker-Planck simulations of proton-beam heating with $E_C$ from 25\,keV to 3\,MeV. They found that low- to moderate-energy proton beams ($E_C$ $\approx$ 50--250\,keV) can drive photospheric perturbations and acoustic transients comparable to observed sunquakes, while higher-energy beams reproduce the continuum enhancements near the Fe\,I 6173\,\AA~line seen in WLF kernels. These results show that deeply penetrating particles can, under certain conditions, reproduce observed photospheric and helioseismic responses that are difficult to obtain in simplified electron-beam models. Nevertheless, the models remain inherently one-dimensional and assume idealized magnetic-loop geometries, precluding investigation of multi-threaded or inclined-field effects.

Although these advances in one-dimensional RHD modeling have significantly improved our understanding of flare energy deposition, they also highlight key limitations that hinder direct comparison with high-resolution observations. In particular, the inherently stratified geometry of 1D models cannot capture the complex magneto-convective structure, lateral transport, and multi-threaded energy deposition that characterize real flare footpoints. The lack of strong-field magnetic topology, fine-scale thermal structure, and multi-dimensional radiative coupling may substantially alter both the depth and efficiency of beam-driven heating. These considerations motivate extending flare-beam modeling to fully three-dimensional radiative-MHD environments with physically self-consistent pre-flare atmospheres, where the interaction between prescribed particle energy deposition, magnetic structuring, and hydrodynamic response can be studied within a self-consistent three-dimensional atmospheric framework and compared directly with Fe\,I 6173\,\AA~and continuum signatures observed by HMI. At the same time, extending to three dimensions introduces new modeling choices and assumptions, particularly regarding magnetic geometry and particle transport, which must be carefully assessed. In the present work, electron beams are represented by an imposed thick-target heating prescription applied vertically through the atmosphere rather than through self-consistent particle transport along magnetic field lines. This approach isolates the atmospheric response to beam energy deposition while omitting field-aligned transport effects considered in more self-consistent particle-transport models.

In this study, we perform three-dimensional StellarBox simulations of electron-beam heating using parameters consistent with the F-CHROMA RADYN grid: total energy flux density $10^{12}$ erg\,s$^{-1}$\,cm$^{-2}$, spectral index $\delta$ = 3, and $E_C$ = 10--25 keV, and synthesize Fe\,I 6173\,\AA~Stokes profiles using the RH 1.5D code \citep{rybicki1991RHI,rybicki1992RHII,uitenbroek2001RHPRD,pereira2015RH15D} for direct comparison with HMI observations. By contrasting the three-dimensional StellarBox results with one-dimensional RADYN electron and proton models under comparable beam parameters, we assess how magnetic and thermal structuring, multidimensional hydrodynamic transport, and radiative coupling influence the atmospheric response to flare energy deposition. Our goal is to determine whether three-dimensional radiative-MHD models can reproduce the depth and magnitude of photospheric heating implied by white-light and helioseismic observations and to identify the physical mechanisms responsible for the most energetic flare kernels.

\section{Observational Constraints}

The Helioseismic and Magnetic Imager obtains narrow-band filtergrams at six wavelength positions across the Fe\,I 6173\,\AA~line in multiple polarizations and produces full Stokes products on standard observing cadences (90\,s for linear polarization sequences and 45\,s for circular polarization). Because the measurement sequence samples the line at discrete filtergram positions over tens of seconds, rapid line shape changes during flares may be imperfectly sampled and must be interpreted with care. For the identification of line-core emission, we follow the core-to-wing (CtW) metric computed from filtergram channels by sampling $\pm$34 mÅ (core) and $\pm$170 mÅ (wing). Although CtW is sensitive to bandpass drifts and complex line shapes, it reliably locates instances where the line core brightens relative to the line wings. In \cite{2025ApJ...988...74G}, several WLFs observed by HMI between 2016 and 2024 that exhibited brief instances of Fe\,I 6173\,\AA~line-core emission were analyzed. These events are spatially compact and short-lived, typically localized to small kernels above sunspot umbrae or umbra/penumbra boundaries and typically last for only a single 90\,s Stokes frame. The most compelling example analyzed in \cite{2025ApJ...988...74G} was the 2017-09-06 X9.3 flare, which displayed multiple umbral emission kernels with CtW ratios approaching about 1.8 and local continuum increases by factors up to 4 in the emission kernels. In the analyzed events, line-core emission always coincides with peaks in hard X-ray and with local maxima in the time derivative of soft X-ray flux, and post-flare analysis often reveals lasting changes in the Stokes Q, U, and/or V profiles at the kernel locations, indicating permanent magnetic field reconfigurations. These observational characteristics constrain plausible heating mechanisms: models must reproduce short duration and small spatial extent, be temporally associated with particle precipitation proxies (e.g., hard X-ray emission), and, in some cases, deliver sufficient impulse to excite helioseismic waves.

\section{Numerical Methods}
\subsection{StellarBox Radiative MHD Simulations}

\begin{figure*}
	\centering
	\includegraphics[width=0.8\textwidth]{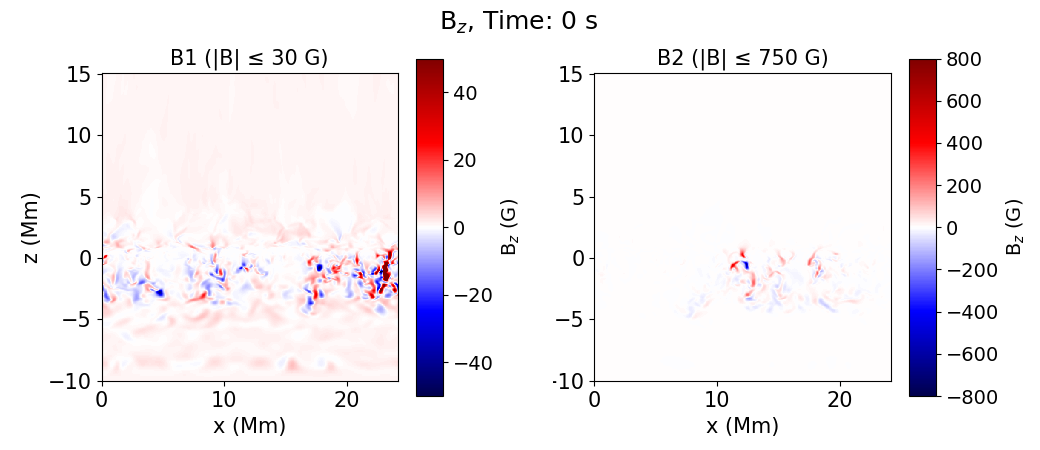}
	\caption{Vertical x-z slices of $Bz$ for the B1 and B2 models before the introduction of the electron beam (t = 0\,s). $|B| \leq 60\,\text{G}$ within 2.5\,Mm of the impact location for the B1 model, and $|B| \leq 750\,\text{G}$ within 2.5\,Mm of the impact location for the B2 model. The slices are taken through the y-coordinate at beam center.}
    \label{B1_B2}
	\vspace{0.1cm}
\end{figure*}

The three-dimensional radiative MHD simulations were performed with the StellarBox code \citep{2015arXiv150707999W}. Simulation domains extend vertically from z = -10\,Mm (subsurface) to z = +15\,Mm (corona) over a 24.2\,$\times$\,24.2\,Mm$^2$ horizontal area using 1\,s time steps. The computational grid consists of 400 points in x and y, with a horizontal grid spacing near 60.7\,km. The vertical grid consists of 400 points and varies in resolution; higher resolutions are used near z = 0\,Mm with spacing in the range of 23.9--157.3\,km. StellarBox solves the compressible MHD equations with multi-group radiative transfer based on opacity binning, together with magnetic-field evolution and convective motions \citep{2015arXiv150707999W}. The radiative transfer treatment uses the Feautrier method with bidirectional ray tracing and opacity-binned frequency groups \citep{2015arXiv150707999W}. The resulting thermodynamic and magnetic structure evolves self-consistently with the MHD equations, providing a structured three-dimensional pre-flare atmosphere into which the prescribed particle beam is introduced. Two representative photospheric magnetic configurations were used (shown in Figure \ref{B1_B2}). The first configuration, denoted B1, centers the beam over a granule with photospheric $|B|$ $\leq$ 60\,G within 2.5\,Mm of the beam impact location. The second configuration, denoted B2, centers the beam over an intergranular lane with photospheric $|B|$ $\leq$ 750\,G within 2.5\,Mm of the beam impact location. The magnetic field in both configurations consists of quiet-Sun magneto-convective structures with mixed-polarity small-scale fields rather than organized coronal loops or strong sunspot fields. The B1 and B2 cases probe how local magneto-convective structure and associated variations in atmospheric density and field strength influence the beam-driven hydrodynamic response. In the absence of a modeled flare reconnection or particle acceleration process, the simulations represent a physically self-consistent pre-flare atmosphere into which flare energy is introduced in a controlled and idealized manner.

\subsection{Particle-beam Implementation}
In this work, electron beams are prescribed rather than self-consistently generated by magnetic reconnection. For simplicity, beams are injected vertically rather than propagated along magnetic field lines. Although the beam is imposed vertically, the local energy deposition depends on the column mass distribution, which is structured by the magnetic field. As a result, energy deposition can occur preferentially in lower-density structures shaped by the magnetic field, without explicitly prescribing field-aligned particle transport. The beam energy flux density is introduced at the top boundary of the computational domain and deposited downward according to local column mass density following the thick-target formulation. This choice allows us to isolate the role of atmospheric stratification and three-dimensional hydrodynamic response, but it neglects field-aligned transport, magnetic mirroring, and return-current effects that may influence energy deposition in real flares. Particle beams were imposed along the vertical (z) direction with a Gaussian profile with a half-width at half maximum (HWHM) of 500\,km (FWHM = 1000\,km, $\sigma \approx$ 425\,km) and a symmetric triangular temporal profile of 20\,s duration. The implementation of collisional energy deposition follows the classical thick-target formalism originally developed by \cite{1972SvA....16..273S} and later extended by \cite{Livshits1981} for calculating the heating rate and ionization effects produced by non-thermal electrons penetrating the solar atmosphere. This formalism is incorporated into the StellarBox framework to compute the local energy-deposition profile and resulting thermal response. While the analytic energy-loss prescription is conceptually similar to that used in RADYN, the present implementation applies the heating directly within the three-dimensional MHD framework without solving a Fokker–Planck transport equation.

A suite of electron beam simulation cases was carried out with a total energy flux density of $1\times10^{12}$ erg\,s$^{-1}$\,cm$^{-2}$ and a power-law spectral index $\delta$ = 3. $E_C$ values of 10, 15, 20, and 25\,keV were chosen to examine the trade-offs between number flux (favoring lower $E_C$) and penetration depth (favoring higher $E_C$). The resulting heating rates were applied to the evolving three-dimensional StellarBox atmosphere at each time step. Proton and mixed-species beams were not included in the present simulations but are planned for future work.

\subsection{RH 1.5D Spectral Synthesis}

For spectral diagnostics, the RH 1.5D radiative transfer code \citep{pereira2015RH15D} was used to synthesize Fe\,I 6173\,\AA~line profiles from vertical columns extracted from the StellarBox atmospheres in a 3.7 $\times$ 3.7\,Mm$^2$ patch centered on the beam impact. RH 1.5D solves the non-local thermodynamic equilibrium (NLTE) radiative transfer problem per column, including Zeeman polarization, and computes Stokes I, Q, U, and V. Temperature, hydrogen populations, magnetic field, vertical velocity, and the height grid are taken directly from the StellarBox model output, and microturbulence was set to 0\,m\,s$^{-1}$ to isolate thermal and bulk velocity effects. The 1.5D approach for calculating the spectral line preserves column stratifications while neglecting fully three-dimensional radiative coupling. To facilitate direct comparison with SDO/HMI filtergram observations, spatially averaged line profiles were computed over a $305 \times 305$\,km$^2$ region centered on the beam impact location. This area is comparable to the spatial sampling of a single HMI filtergram pixel ($0.5" \approx 360 \text{\,km}$ at disk center). While this averaging suppresses sub-pixel extremes, it provides an observationally motivated representation of unresolved flare kernels and allows direct assessment of whether Fe I line-core emission would be detectable by HMI.

\section{Results}
\subsection{Atmospheric Heating and Hydrodynamic Response}

\begin{figure*}
	\centering
	\includegraphics[width=0.8\textwidth]{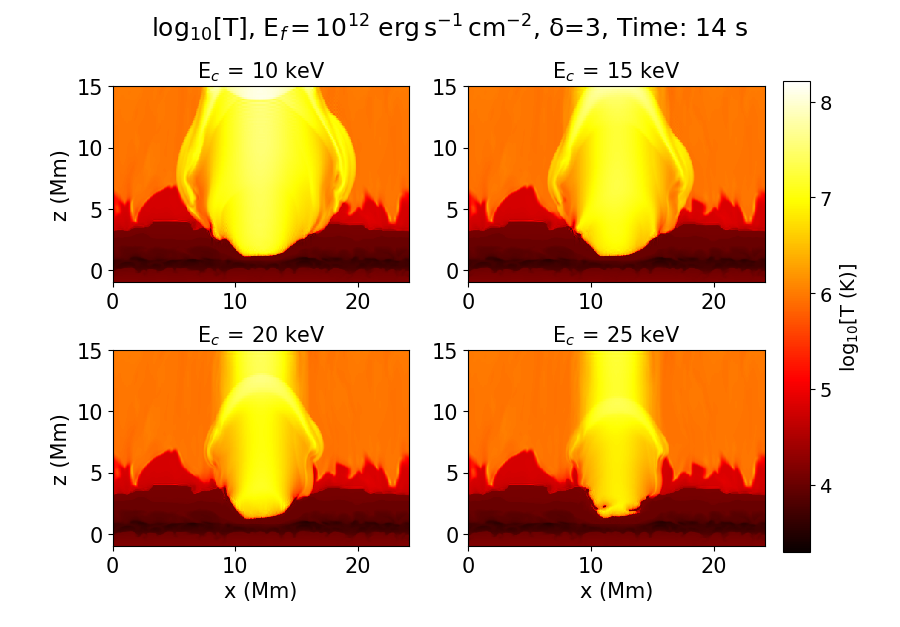}
	\caption{Vertical x-z slices of $\log_{10}$ temperature ($T$) for the B1 model for the $E_C$ = 10, 15, 20, and 25\,keV beams at t = 14\,s. The slices are taken through the y-coordinate at beam center.}
    \label{B1_LogT}
	\vspace{0.1cm}
\end{figure*}

\begin{figure*}
	\centering
	\includegraphics[width=0.8\textwidth]{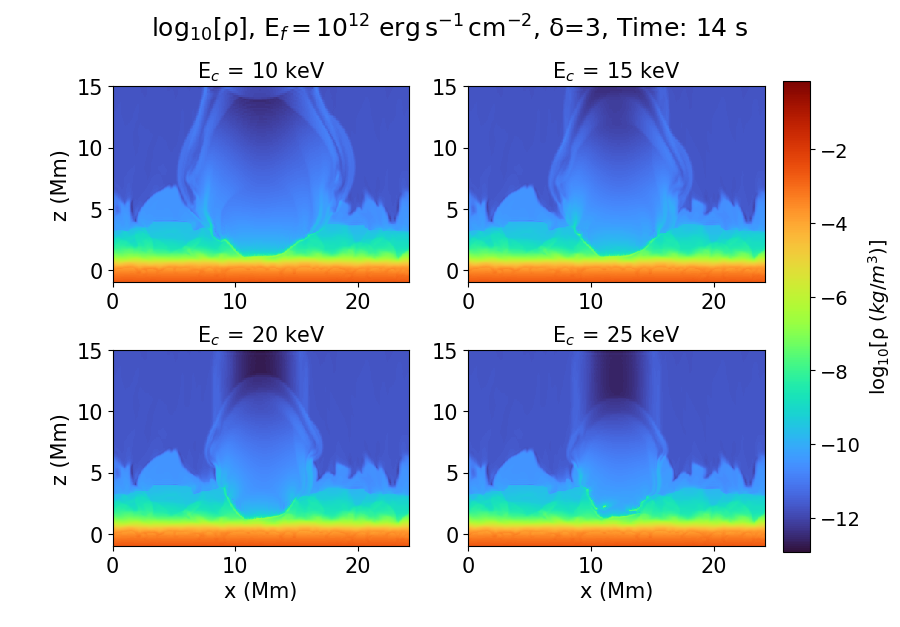}
	\caption{Vertical x-z slices of $\log_{10}$ mass density ($\rho$) for the B1 model for the $E_C$ = 10, 15, 20, and 25\,keV beams at t = 14\,s. The slices are taken through the y-coordinate at beam center.}
    \label{B1_Logrho}
	\vspace{0.1cm}
\end{figure*}

Across the electron-beam parameter set, heating remains concentrated in the upper chromosphere and above $z \approx 1.25$~Mm in both magnetic configurations. Lower $E_C$ beams, with their larger number fluxes, deposit energy higher in the atmosphere and produce a more rapid hydrodynamic response, whereas higher $E_C$ beams penetrate to lower heights (smaller $z$) but yield weaker upper-chromospheric heating (see Figures \ref{B1_LogT} and \ref{B1_Logrho}). This qualitative dependence of heating height and response time on $E_C$ is also expected in one-dimensional thick-target models, but the three-dimensional simulations reveal a more spatially fragmented and laterally extended response due to the structured atmosphere.

\begin{figure*}
	\centering
	\includegraphics[width=1\textwidth]{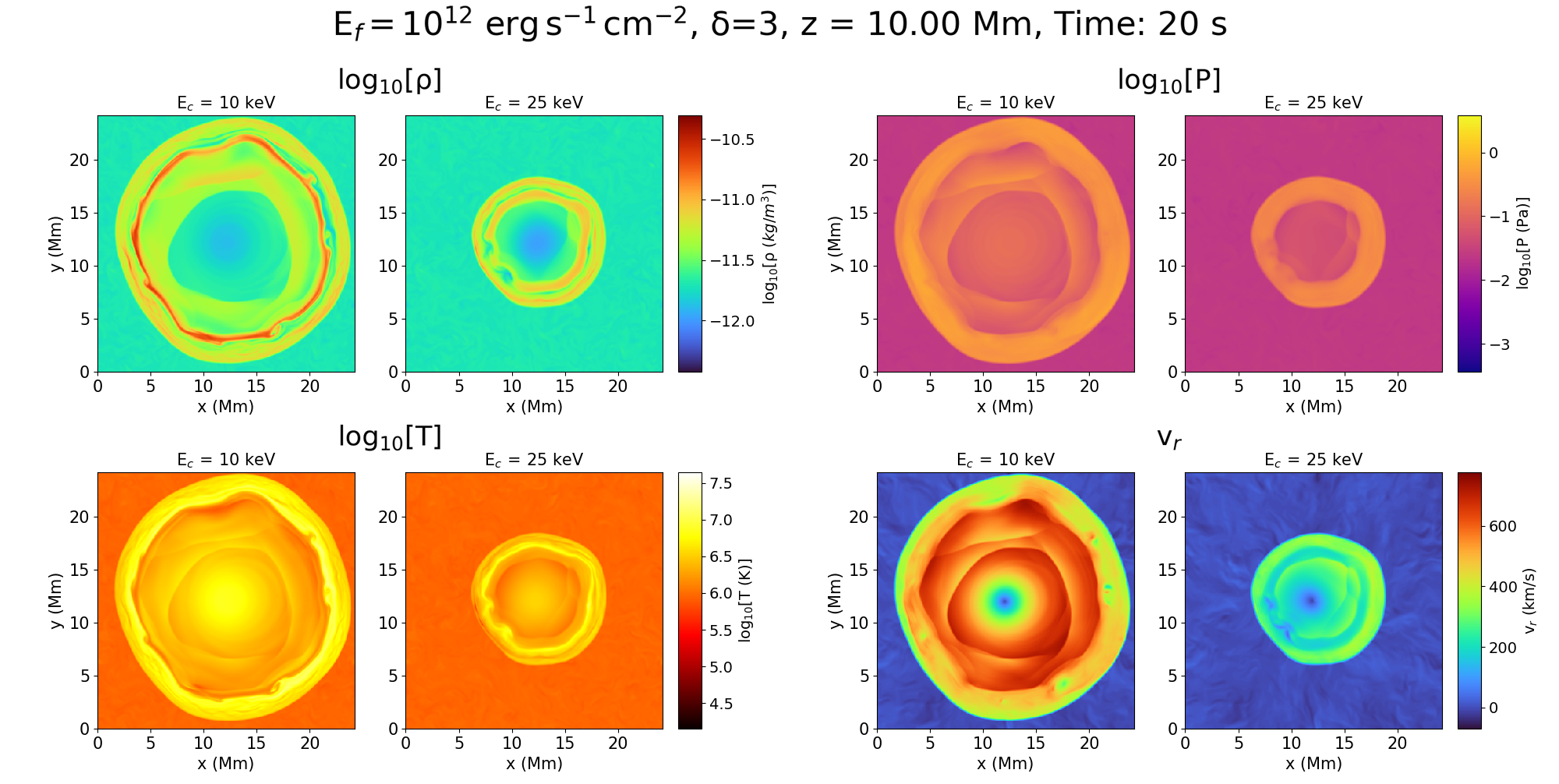}
	\caption{Horizontal x-y Corona slices at z = 10.00 Mm for $\log_{10}$ mass density ($\rho$), $\log_{10}$ pressure ($P$), $\log_{10}$ temperature ($T$), and local horizontal radial plasma velocity ($v_r$) for the B1 model for the $E_C$ = 10 and 25\,keV beams at t = 20\,s.}
    \label{z_10.00Mm}
	\vspace{0.1cm}
\end{figure*}

\begin{figure*}
	\centering
	\includegraphics[width=1\textwidth]{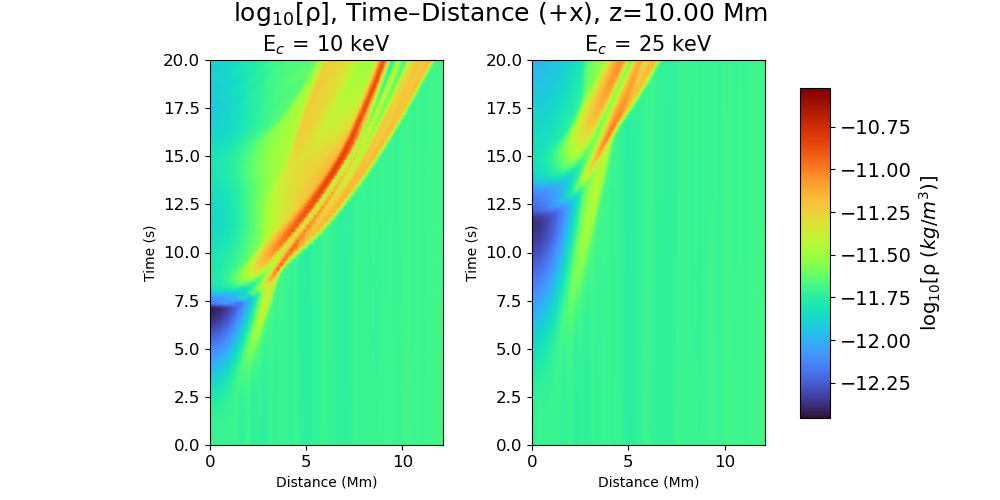}
	\caption{Time-distance plots of $\log_{10}$ mass density ($\rho$) at z = 10.00 Mm for the B1 model for the $E_C$ = 10 and 25\,keV beams. Distance is measured along the +x direction where Distance = 0 corresponds to the coordinates at beam center.}
    \label{time_distance_rho}
	\vspace{0.1cm}
\end{figure*}

\begin{figure*}
	\centering
	\includegraphics[width=1\textwidth]{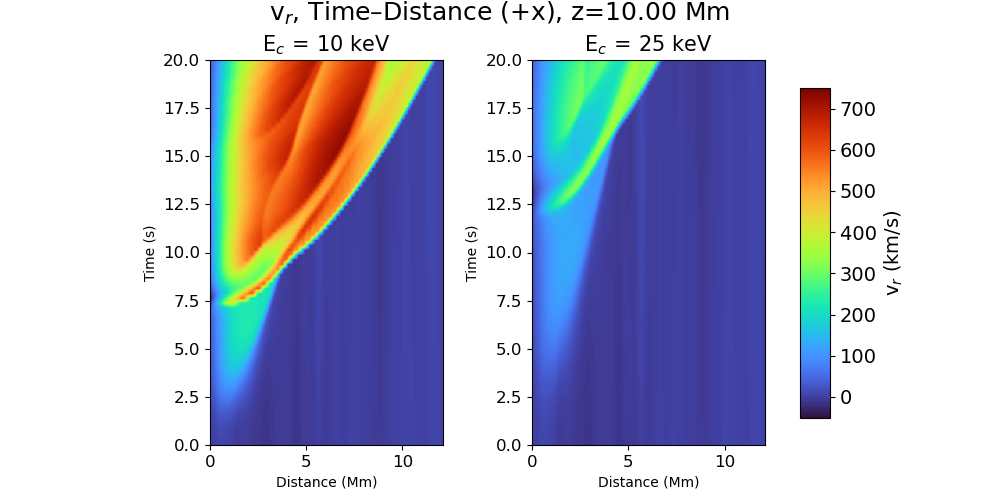}
	\caption{Time-distance plots of local horizontal radial plasma velocity $v_r$ at z = 10.00 Mm for the B1 model for the $E_C$ = 10 and 25\,keV beams. Distance is measured along the +x direction where Distance = 0 corresponds to the coordinates at beam center.}
    \label{time_distance_vr}
	\vspace{0.1cm}
\end{figure*}

In the B1 simulation, the onset of beam heating generates a high-pressure disturbance visible within $\sim 5$\,s of beam initiation. In the 10\,keV case, this produces a prominent upward-propagating shock earliest, with the 15--25\,keV models showing comparable fronts appearing progressively later ($\sim$ 6--9\,s). Multiple upward-moving compression fronts develop in the corona, three to four in the 10\,keV run and two in the 25\,keV run, readily identifiable in the radial-velocity and density signatures at $z = 10$\,Mm (Figures \ref{z_10.00Mm}, \ref{time_distance_rho}, and \ref{time_distance_vr}). At this height, each model exhibits two sharp outer density rings and, in the 10\,keV case, an additional, more diffuse inner ring. The 25\,keV simulation shows two outer rings of similar density, while the 10\,keV case displays a second ring of noticeably higher density than the first. In both models, the highest temperature and pressure occur between the first two rings. Time-distance diagrams of the local horizontal radial plasma velocity reveal an initial, weaker disturbance propagating outward with a front-propagation speed of $\sim$ 350\,km\,s$^{-1}$ and $\sim$ 230\,km\,s$^{-1}$ in the 10 and 25 keV runs, respectively. The local horizontal radial plasma velocities immediately behind these disturbances reach maximum values of $\sim$ 230\,km\,s$^{-1}$ and $\sim$ 130\,km\,s$^{-1}$ in the 10\,keV and 25\,keV runs respectively with almost no vertical velocity contribution. A much stronger shock emerges later at $\sim$ 7.5\,s in the 10\,keV case with an initial shock-front velocity of $\sim$ 1070\,km\,s$^{-1}$, and at $\sim$ 13\,s in the 25\,keV case with an initial shock-front velocity of $\sim$ 675\,km\,s$^{-1}$. The local horizontal radial plasma velocities behind these stronger shocks range between $\sim$ 600--800\,km\,s$^{-1}$ and $\sim$ 300--400\,km\,s$^{-1}$ in the 10\,keV and 25\,keV runs respectively, again with almost no vertical velocity contribution. Subsequent shock fronts travel more slowly, though at speeds comparable to the leading fronts, and remain significantly faster than the initial weak disturbance. However, the local plasma velocity behind these subsequent shocks increases with a much greater contribution from the vertical component. In the 10\,keV case, the vertical velocity reaches $\sim$ 1600--2000\,km\,s$^{-1}$ near the center of the beam between t = 9 and 19\,s. In the 25\,keV case, the vertical velocity reaches $\sim$ 800--1000\,km\,s$^{-1}$ near the center of the beam between t = 14 and 19\,s. 

\begin{figure*}
	\centering
	\includegraphics[width=1\textwidth]{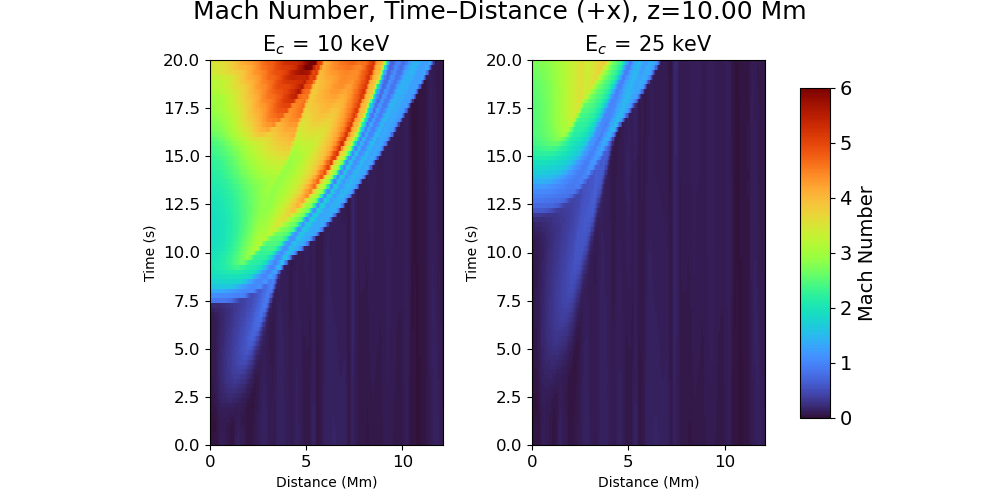}
	\caption{Time-distance plots of the local flow Mach number, $M=|\mathbf{v}|/c_s$, where $|\mathbf{v}|$ is the full 3D local plasma velocity magnitude and $c_s=(\gamma P/\rho)^{1/2}$ is the local adiabatic sound speed, with $\gamma=5/3$. Distance is measured along the +x direction where Distance = 0 corresponds to the coordinates at beam center.}
    \label{time_distance_M}
	\vspace{0.1cm}
\end{figure*}

To assess whether the large velocities associated with the propagating disturbances correspond to supersonic flows, we calculate the local flow Mach number, $M=|\mathbf{v}|/c_s$, where $c_s=(\gamma P/\rho)^{1/2}$ is the local adiabatic sound speed and $\gamma=5/3$ (Figure~\ref{time_distance_M}). The initial, slower disturbance remains approximately transonic or subsonic, with local flow Mach number $M\lesssim1$ immediately behind the shock front in both the 10 and 25\,keV models. The first fast compression front is followed by local plasma flows with $M \approx$ 1.4--1.5. Subsequent shocks are associated with substantially more supersonic post-front flows, with the 10\,keV model reaching $M\approx$ 5--5.2 behind the second fast shock and $M\approx$ 5.5--6 behind the third. The corresponding Mach numbers in the 25\,keV model are lower, reaching $M\approx$ 2.3--2.4 behind the second fast shock and $M\approx$ 3.6--3.8 behind the third. Thus, the high velocities in the later disturbances cannot be attributed solely to the elevated adiabatic sound speed of the hot corona; the plasma flows immediately associated with these disturbances are locally supersonic. The larger Mach numbers in the 10\,keV case are consistent with its earlier and more energetic hydrodynamic response, while the higher-energy 25\,keV beam produces weaker but still strongly supersonic subsequent shocks.

Although local plasma velocities in the corona reach $\sim$ 1000--2000\,km\,s$^{-1}$, these large velocities are strongly localized to the upper atmosphere and decrease rapidly with depth. In all models, shock-associated velocities remain below $\sim$ 100\,km\,s$^{-1}$ below $z\approx1.25$\,Mm and under $\sim$ 10\,km\,s$^{-1}$ below $z\approx1$\,Mm. The Mach number analysis demonstrates that the large coronal velocities correspond to genuinely supersonic disturbances, particularly for the later shock fronts, rather than simply reflecting the high adiabatic sound speed of the hot corona. Nevertheless, the rapid decrease in both velocity and shock strength with decreasing height indicates that these disturbances do not transport substantial momentum into the photosphere in the present simulations.

These shocks originate from the rapid, localized overpressure created when the electron beam deposits energy impulsively in the upper chromosphere and lower corona. The resulting steep pressure gradients drive upward and lateral flows that steepen into shocks as they expand into progressively lower-density plasma. Because the local magnetic field in these models is weak in the chromosphere and corona ($\leq$\,10\,G), magnetic forces contribute little to the dynamics, and the disturbances behave primarily as hydrodynamic shocks, with magnetic forces playing a secondary role in their propagation.

\begin{figure*}
	\centering
	\includegraphics[width=0.9\textwidth]{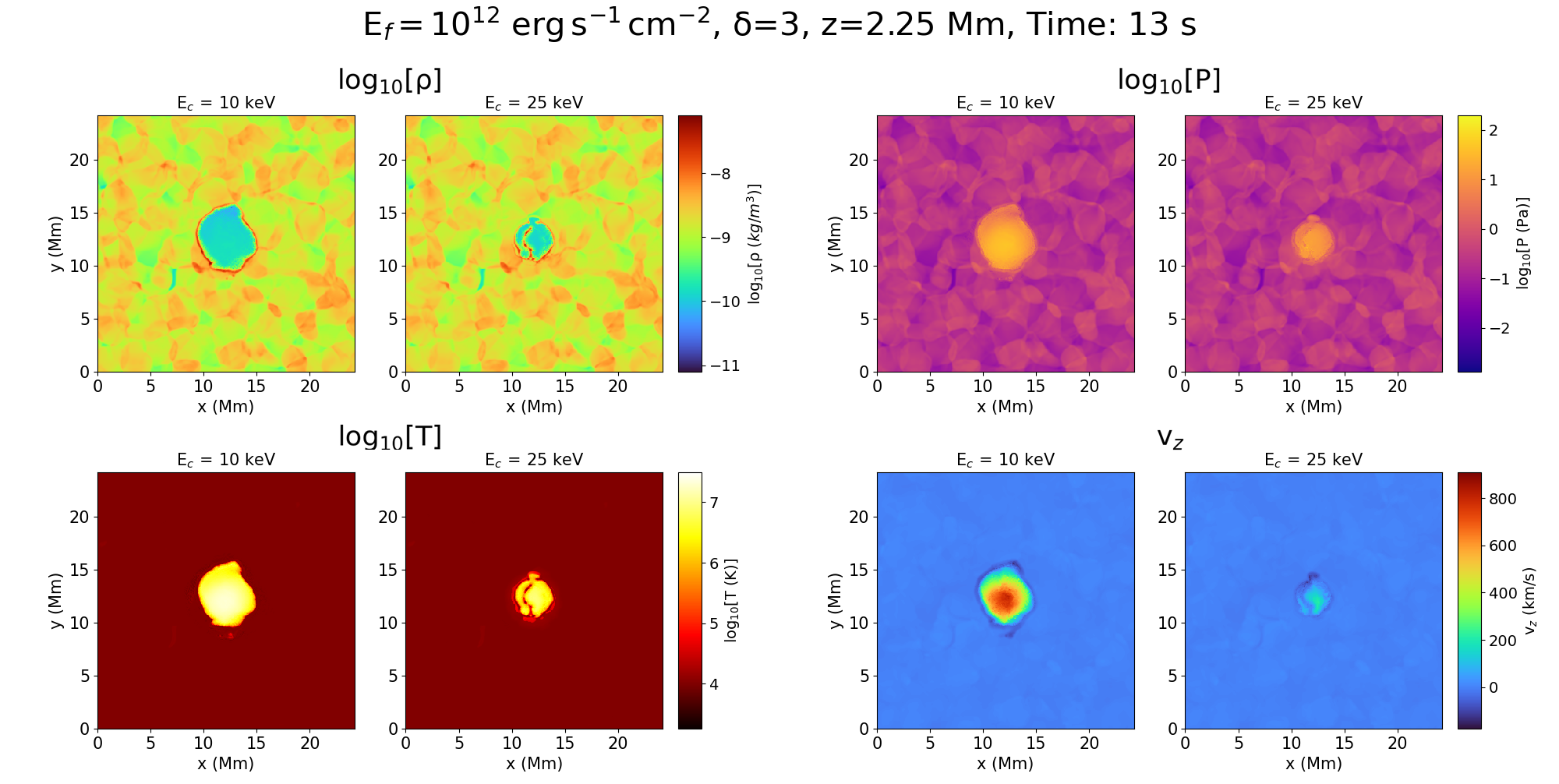}
	\caption{Horizontal x-y Chromosphere slices at z = 2.25 Mm for $\log_{10}(\rho)$, $\log_{10}(P)$, $\log_{10}(T)$, and $v_z$ for the B1 model for the $E_C$ = 10 and 25\,keV beams at t = 13\,s.}
    \label{z_2.25Mm}
	\vspace{0.1cm}
\end{figure*}

At chromospheric heights ($z \approx 2.25$\,Mm), the hydrodynamic response exhibits strong lateral structuring. In the B1 configuration these structures develop above the weak-field granule centered on the beam axis, with lateral spreading into adjacent intergranular lanes. These features are confined to within $\sim$2\,Mm of the beam axis in the horizontal plane and are centered on the beam impact region shown in Figure~\ref{z_2.25Mm}. In the 25\,keV case, chromospheric condensation becomes visible by $\approx$ 8\,s, initially forming a curved, asymmetric condensation near beam center before spreading outward and fragmenting as heating continues. A secondary ring-like condensation forms around the beam perimeter between $\approx$ 10 and 15\,s. In the 10\,keV model, a weaker condensation is faintly visible along the beam perimeter between $\approx$ 6--10\,s. In both models, beam heating produces a central region of reduced density surrounded by a denser annulus tracing the condensation front, accompanied by enhanced pressure near the beam core. Vertical velocities are upward in the central impact region (larger in the 10\,keV model), while small downward velocities appear at the condensation sites along the perimeter. 

In the 25\,keV case, a compact “chromospheric bubble” forms beneath the primary heating site as a cool, dense pocket that expands laterally over time, consistent with the bubble-like structures described by \cite{2020ApJ...894L..21R}. Because the present simulations do not include loop-scale magnetic confinement and the strongest response undergoes substantial lateral expansion, the evolution of the condensation is dominated by pressure-driven compression, radiative losses, and expansion rather than by the loop-top thermal instability typically discussed in confined flare loops. Thermal conduction may also contribute to redistribution of thermal energy, but its relative importance is not isolated in the present analysis. The combination of chromospheric condensation, bubble formation, and multiple upward-propagating shocks reflects the complex interactions between rapid heating, pressure-gradient expansion, and modest magnetic structuring. Despite these highly dynamic effects, the dominant beam-driven heating remains above $z \approx 1.25$\,Mm.

\newpage

\subsection{Comparison with F-CHROMA RADYN 1D Models}
\begin{figure*}[h]
	\centering
	\includegraphics[width=0.7\textwidth]{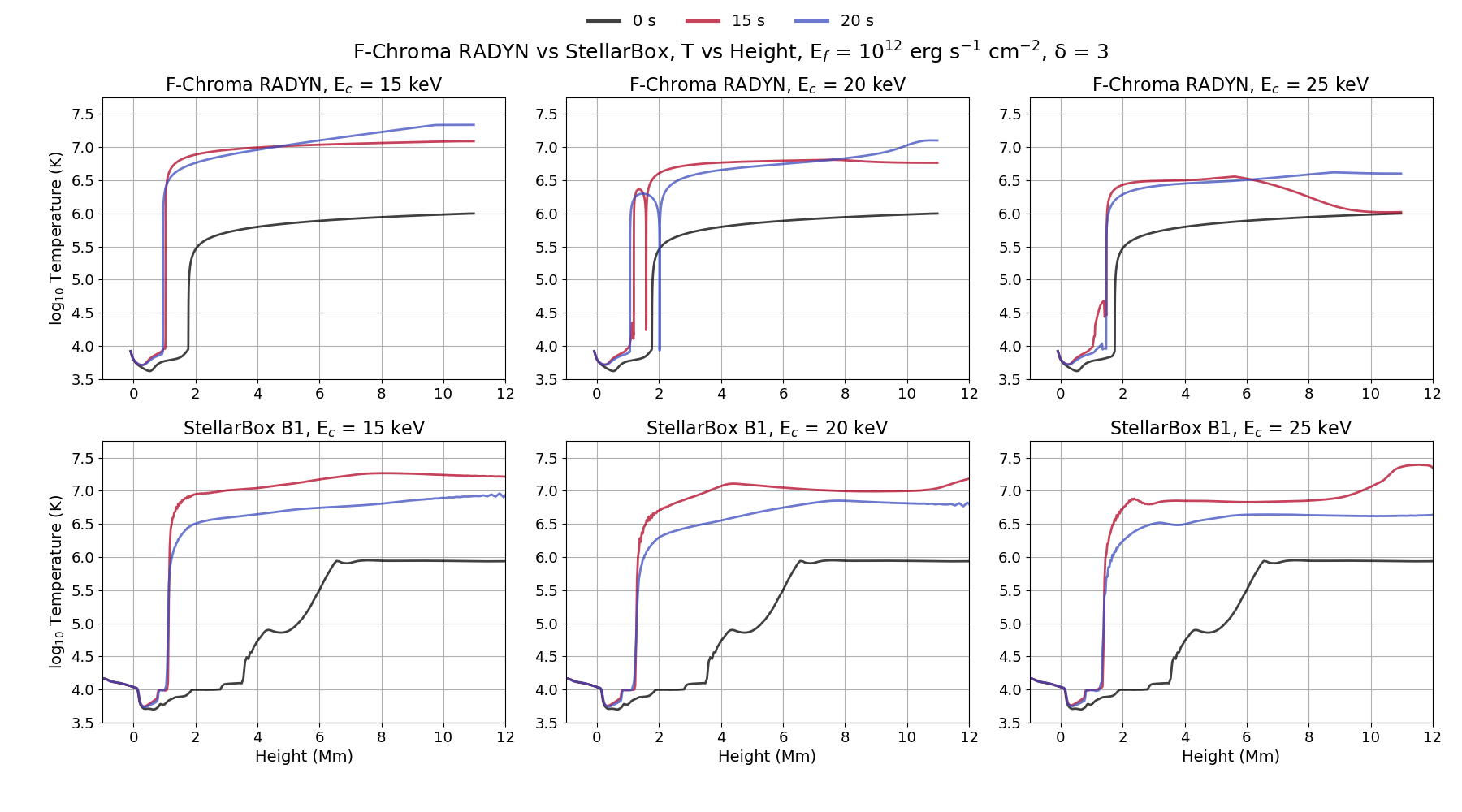}
	\caption{$\log_{10}(T)$ vs height for the F-CHROMA RADYN (top) and B1 StellarBox (bottom) beam models for $E_C$ = 15, 20, and 25\,keV beams at t = 0, 15, and 20\,s. The StellarBox profiles are averaged over a circular region of radius 0.25\,Mm centered on the beam impact. The height range is truncated at z = 12\,Mm to exclude reflections from the upper boundary of the computational domain in the StellarBox model.}
    \label{Stel_vs_RAD_T}
	\vspace{0.1cm}
\end{figure*}

\begin{figure*}
	\centering
	\includegraphics[width=0.7\textwidth]{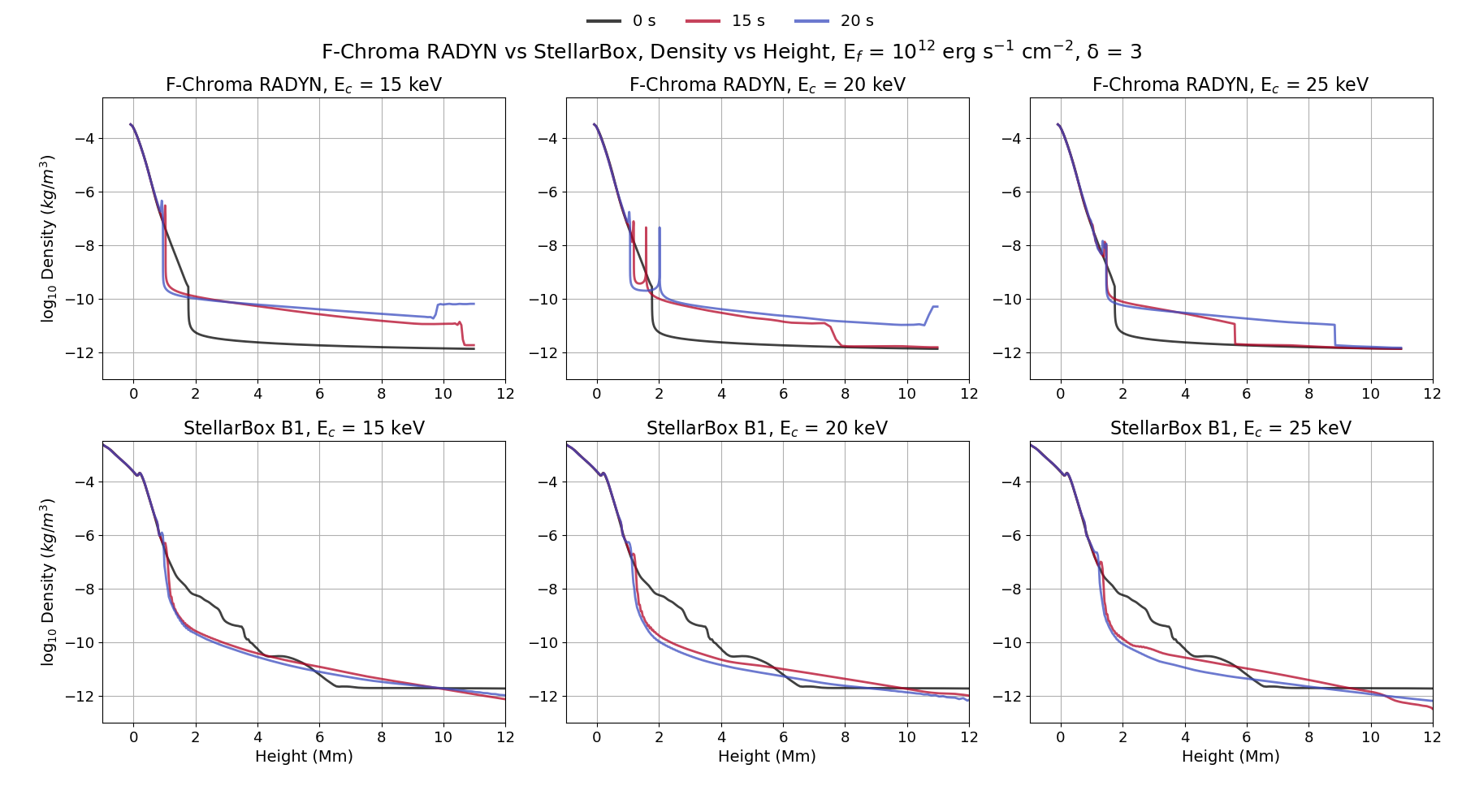}
	\caption{$\log_{10}(\rho)$ vs height for the F-CHROMA RADYN (top) and B1 StellarBox (bottom) beam models for $E_C$ = 15, 20, and 25\,keV beams at t = 0, 15, and 20\,s. The StellarBox profiles are averaged over a circular region of radius 0.25\,Mm centered on the beam impact. The height range is truncated at z = 12\,Mm to exclude reflections from the upper boundary of the computational domain in the StellarBox model.}
    \label{Stel_vs_RAD_rho}
	\vspace{0.1cm}
\end{figure*}

To place the three-dimensional StellarBox results in context, we compared them with one-dimensional F-CHROMA RADYN flare models \citep{Carlsson2023} computed with comparable beam parameters: total energy flux density of $1\times10^{12}$ erg\,s$^{-1}$\,cm$^{-2}$, a power-law spectral index $\delta$ = 3, and a 20\,s triangular temporal profile. Because no 10\,keV RADYN model is available in the F-CHROMA database, comparisons were made for the 15, 20, and 25\,keV cases. The primary variables examined were temperature and column density as a function of height and time, allowing direct comparison of heating depth, coronal response, and cooling behavior between the two model frameworks (Figures \ref{Stel_vs_RAD_T} and \ref{Stel_vs_RAD_rho}). As in the one-dimensional RADYN models, increasing $E_C$ leads to deeper penetration of beam energy and reduced heating in the upper chromosphere, while chromospheric condensation develops beneath the primary heating region. These similarities indicate that the fundamental thick-target energy deposition physics is preserved in the three-dimensional implementation.

Prior to beam injection, the StellarBox atmospheres exhibit substantially greater chromospheric and coronal structural detail in contrast to the relatively smooth RADYN models (Figures \ref{Stel_vs_RAD_T} and \ref{Stel_vs_RAD_rho}). In the RADYN simulations, the transition region appears as an abrupt jump near z $\approx$ 2\,Mm from $\approx 1\times10^4$\,K to $\approx 1\times10^6$\,K, whereas in StellarBox the transition from chromospheric to coronal temperatures is more gradual, extending from z $\approx$ 2 to 6\,Mm with temperature increasing approximately linearly in $\log_{10}(T)$ from $\approx 1\times10^4$\,K to $\approx 1\times10^6$\,K. During beam heating, temperature increases occur more rapidly at all heights in StellarBox than in the corresponding RADYN runs. The maximum coronal temperatures reached in RADYN are $\approx 3\times10^6$ to $1\times10^7$\,K, decreasing with increasing $E_C$, while StellarBox produces higher coronal peaks of $\approx 1\times10^7$ to $2\times10^7$\,K that follow the same trend with $E_C$ (Figure \ref{Stel_vs_RAD_T}).

Another key difference is the post-heating evolution. Faster decay occurs in the three-dimensional StellarBox simulations, partly due to lateral expansion and expansion-driven cooling processes absent in the 1D RADYN geometry. In the StellarBox model, the atmosphere begins to cool during the decay phase of the beam, roughly after the 15\,s mark, with coronal temperatures decreasing by between 0.2 to 0.8\,MK between t $\approx$ 15 and 20\,s depending on the beam $E_C$ (Figure~\ref{Stel_vs_RAD_T}), and reaches near pre-beam values within roughly 10\,s after beam termination. In comparison, the coronal layers in the RADYN models remain elevated well beyond the end of heating. Additionally, the fine-scale temperature structure present in the StellarBox chromosphere and transition region before beam impact becomes smoothed during the impulsive phase. Finally, a small chromospheric bubble is visible in the 20\,keV RADYN simulation similar to the one visible in the 25\,keV StellarBox model, manifesting as a rising localized cool region near z $\approx$ 1.75 -- 2\,Mm  consistent with bubble-like features previously reported in flare simulations \citep{2020ApJ...894L..21R}. These structures may arise from similar pressure-driven dynamics despite the different dimensionality of the models.

\newpage
\subsection{Fe\,I 6173\,\AA~Synthesis}
\begin{figure*}[h]
	\centering
	\includegraphics[width=0.7\textwidth]{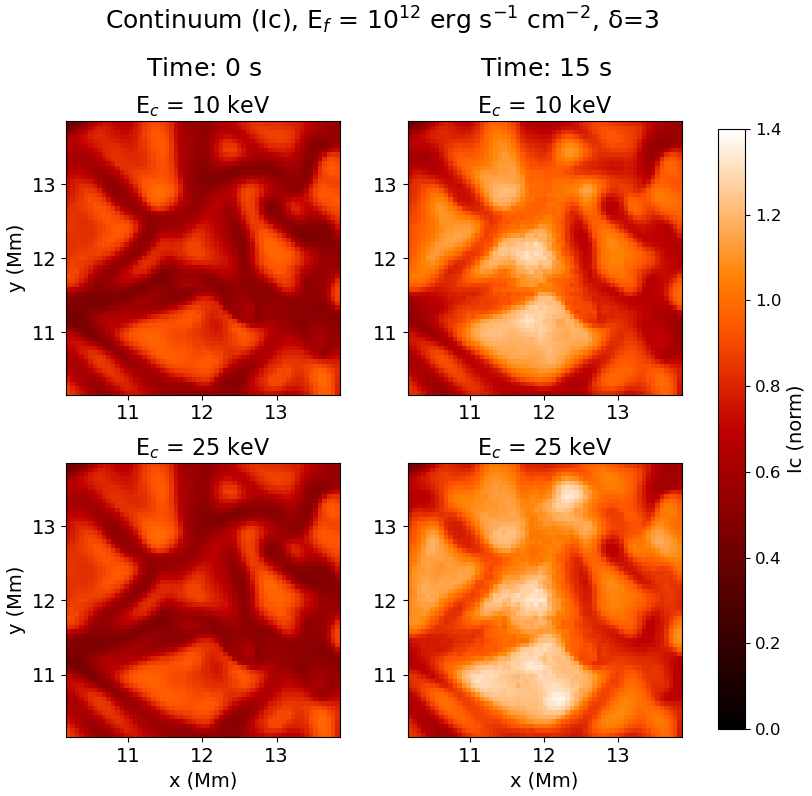}
	\caption{Continuum intensity near Fe\,I 6173\,\AA~for the B1 model for the $E_C$ = 10 and 25\,keV beams at t = 0, 15\,s. Continuum intensity is normalized to the largest value at t = 0\,s.}
    \label{Continuum}
	\vspace{0.1cm}
\end{figure*}

\begin{figure*}[h]
	\centering
	\includegraphics[width=0.7\textwidth]{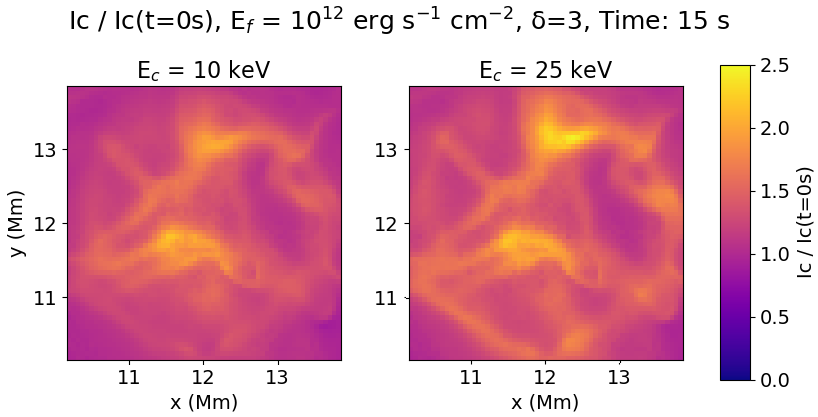}
	\caption{Continuum intensity normalized to t = 0\,s near Fe\,I 6173\,\AA~for the B1 model for the $E_C$ = 10 and 25\,keV beams at t = 15\,s. The maximum achieved values were 2.17 for $E_C$ = 10\,keV and 2.47 for $E_C$ = 25\,keV, with both occurring at t = 15\,s.}
    \label{NormIc}
	\vspace{0.1cm}
\end{figure*}

NLTE line synthesis reveals strong continuum enhancement in all electron-beam runs, with peak continuum intensity increases reaching factors of about 2.17 and 2.47 relative to pre-beam levels for the $E_C$ = 10 and 25 keV beam cases respectively, and with the onset of enhancement occurring slightly earlier for higher $E_C$ runs (Figures \ref{Continuum} and \ref{NormIc}). The normalized continuum plots (Figure \ref{NormIc}) show that this enhancement is spatially structured, with the largest relative increases concentrated in intergranular lanes, reflecting the influence of local photospheric conditions on the continuum response. 

\begin{figure*}[h]
	\centering
	\includegraphics[width=0.7\textwidth]{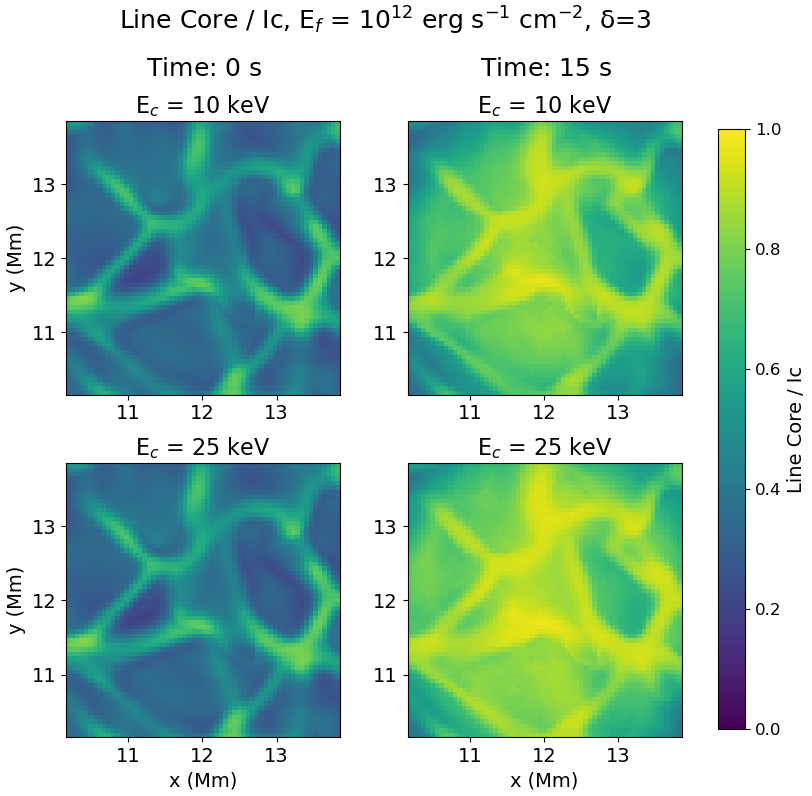}
	\caption{Fe\,I 6173\,\AA~line core for the B1 model for the $E_C$ = 10 and 25\,keV beams at t = 0, 15\,s. Line core is normalized to the local continuum intensity. The maximum achieved values were 0.965 for $E_C$ = 10\,keV and 0.978 for $E_C$ = 25\,keV, with both occurring at t = 15\,s.}
    \label{Line Core}
	\vspace{0.1cm}
\end{figure*}

The Fe\,I 6173\,\AA~line core brightens in all cases, and the core enhancement generally increases with $E_C$, yet the line remains in absorption for all tested $E_C$ values. The peak core intensity does not exceed the continuum at any location, and reaches a maximum of 96.5\% of continuum for $E_C$ = 10\,keV and a maximum of 97.8\% of continuum for $E_C$ = 25\,keV (Figure \ref{Line Core}). While electron beams with the considered parameters reproduce continuum WLF signatures, the Fe\,I 6173\,\AA~line remains in absorption in all cases, indicating that additional factors, such as deeper energy deposition or cooler pre-flare atmospheric conditions, may be required to produce full line-core emission.

Spatially, the line approaches emission most closely above the intergranular lanes (Figures \ref{Continuum} and \ref{Line Core}). These regions are characterized by cooler photospheric temperatures compared to the adjacent granules. This sensitivity suggests that cooler pre-flare stratifications may favor line-core reversal. Although the present models contain quiet-Sun granulation rather than sunspot atmospheres, this behavior is qualitatively consistent with the preferential occurrence of HMI Fe\,I 6173\,\AA~emission kernels in the cooler umbral and umbra-penumbra environments. We note that the present simulations sample quiet-Sun granulation rather than sunspot umbrae or penumbrae, where the magnetic field strength and pre-flare thermal stratification differ substantially and may further favor line-core emission. This supports the interpretation that cooler atmospheric stratifications are more susceptible to line reversal.

\begin{figure*}[h]
	\centering
	\includegraphics[width=0.7\textwidth]{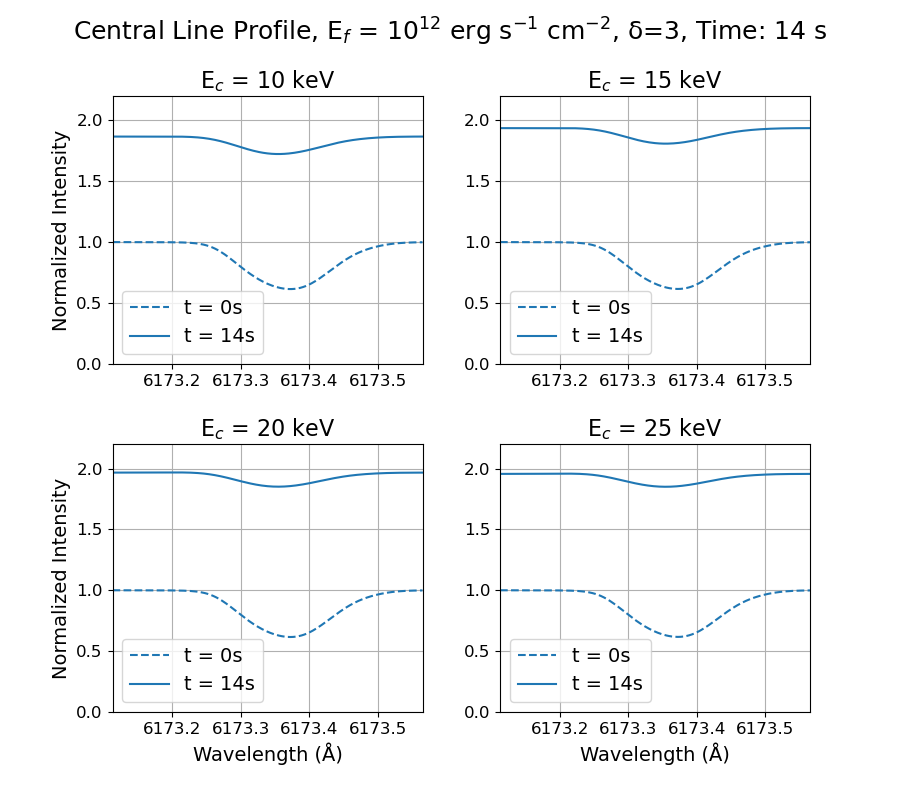}
	\caption{Synthetic Fe\,I 6173\,\AA~line profiles computed for the B1 StellarBox model averaged over the $305\times305$ km$^2$ area centered on the beam impact. This area is comparable to the spatial sampling of a single HMI filtergram pixel ($0.5" \approx 360 \text{\,km}$ at disk center). Profiles are shown for the $E_C$ = 10, 15, 20, and 25\,keV beams at t = 0 and 14\,s}
    \label{6173A_lines}
	\vspace{0.1cm}
\end{figure*}

Finally, the synthesized spatially averaged line profiles (Figure~\ref{6173A_lines}) show that although all $E_C$ values produce substantial line-core brightening, none achieve a transition to full emission at any time. The spatial averaging is intended to represent the signal that would be measured within a single SDO/HMI filtergram pixel, whose spatial sampling ($0.5" \approx 360 \text{\,km}$ at disk center) is comparable to the averaging scale used here, and thus provides an observationally motivated measure of detectability. Under this constraint, the beam parameters studied here do not yield the deep photospheric heating needed to reproduce the Fe\,I 6173\,\AA~emission observed in strong white-light kernels.


\vspace{1.5cm}

\section{Discussion}
The three-dimensional radiative MHD simulations presented here reveal that electron beams with parameters representative of strong flare electron distributions can drive substantial upper-chromospheric heating and reproduce strong continuum enhancements, while placing constraints on the conditions required to produce Fe\,I 6173\,\AA~line-core emission. In all cases, beam energy deposition and the resulting heating remain concentrated above $z \approx 1.25$\,Mm (Figures \ref{B1_LogT} and \ref{B1_Logrho}), and the resulting shocks, condensations, and chromospheric bubbles evolve on short (5--20\,s) timescales without appreciable penetration into the photosphere. Although the atmospheric response is highly structured and strongly time-dependent, the depth of energy deposition limits the ability of these disturbances to produce the compact, high-contrast continuum kernels seen in observations. We emphasize that the present simulations employ relatively weak quiet-Sun magnetic fields and a vertically imposed beam geometry rather than field-aligned particle transport. These assumptions simplify the implementation but may alter both the spatial distribution and depth of energy deposition compared to flare loops rooted in strong-field active-region environments.

Comparison with the one-dimensional F-CHROMA RADYN flare models clarifies the role of multidimensional geometry and self-consistent atmospheric structure. Despite using identical energy flux densities, spectral indices, and temporal profiles, the StellarBox atmospheres exhibit significantly greater chromospheric and coronal structuring prior to beam injection (Figures \ref{Stel_vs_RAD_T} and \ref{Stel_vs_RAD_rho}), reflecting magneto-convective dynamics absent in 1D equilibria. This fine structuring modulates beam energy deposition, accelerates the formation of shocks, and enables more vigorous lateral expansion. Consequently, the StellarBox simulations produce faster and more spatially extended temperature increases, along with higher peak coronal temperatures ($\approx 1\times10^7$ to $2\times10^7$\,K) than their RADYN counterparts ($\approx 3\times10^6$ to $1\times10^7$\,K) (Figure \ref{Stel_vs_RAD_T}). These differences demonstrate how multidimensional hydrodynamic transport and pre-existing thermal and density structuring shape the flare response even when beam parameters are held fixed.

A striking contrast with 1D models is the rapid cooling observed in StellarBox. Temperatures return to near pre-flare values within about 10\,s of beam termination (Figure \ref{Stel_vs_RAD_T}), consistent with rapid cooling associated in part with lateral expansion, with radiative losses also contributing. In particular, lateral expansion of heated plasma enables expansion-driven cooling and redistribution of energy into surrounding regions, reducing local energy confinement compared to 1D geometries. In contrast, the 1D RADYN models retain hot coronal plasma long after the beam ceases, highlighting differences in energy confinement that stem from the absence of lateral expansion and multidimensional redistribution of heated plasma, rather than fundamentally different radiative loss mechanisms. These distinctions demonstrate the need for multi-dimensional modeling when evaluating flare energy transport and post-impulsive cooling.

The synthetic Fe\,I 6173\,\AA~profiles provide further insight into the observational consequences of these hydrodynamic processes. All electron-beam cases exhibit strong continuum brightening consistent with continuum enhancements observed during strong X-class white-light flares (Figures \ref{Continuum}--\ref{6173A_lines}), but none achieve full line-core reversal. Instead, the line brightens substantially while remaining in absorption, with core intensities reaching at most $\approx$ 97--98\% of the continuum. The line approaches emission most closely above the intergranular lanes, where cooler pre-flare photospheric temperatures lower the continuum intensity and thus enhance the relative line-core contrast. This behavior is consistent with HMI observations \citep{2025ApJ...988...74G}, where Fe\,I core emission kernels are concentrated in umbrae and umbra-penumbra boundaries. We note that the present simulations represent quiet-Sun granulation rather than strong-field sunspot atmospheres, where both the magnetic structure and cooler pre-flare stratification may further enhance the likelihood of line-core emission. Thus, the susceptibility of Fe\,I 6173\,\AA~to reversal appears to depend on both the heating depth and the pre-flare thermal stratification.

The shock and condensation dynamics visible in the simulations highlight important implications for helioseismic responses. Multiple upward shock fronts, lateral expansions, and chromospheric condensations introduce significant momentum fluxes into the overlying plasma. Under certain conditions, these motions may contribute to the excitation of downward-propagating acoustic waves, a key requirement for sunquake generation. However, because the impulsive heating remains confined to heights above the low chromosphere, these electron-beam-driven dynamics may be insufficient to reproduce the strongest helioseismic impulses under the atmospheric and magnetic conditions explored here. In particular, the absence of significant overpressure or momentum transfer near the photosphere suggests that an additional, deeper-reaching energy carrier is required. 

The coronal velocities reaching $\sim$1000--2000\,km\,s$^{-1}$ are substantially higher than the several-hundred km\,s$^{-1}$ upflows commonly measured in hot flare plasma. For example, \cite{2013ApJ...776L..11I} reported hot ($\sim$30\,MK) flows exceeding $500$\,km\,s$^{-1}$ above the arcade during an X-class flare. The local flow Mach number analysis indicates that these velocities cannot be attributed solely to the high adiabatic sound speed of the heated corona: the later disturbances reach $M\approx$ 5--6 in the 10\,keV case and $M\approx$ 2--4 in the 25\,keV case. The high velocities therefore represent genuinely supersonic plasma motions associated with the strong pressure-driven shocks produced by the highly impulsive and spatially concentrated beam heating. Their occurrence should nevertheless be interpreted cautiously, since the simulations employ an idealized vertically imposed beam and weak coronal magnetic fields, and the resulting hydrodynamic response may differ from that of a magnetically confined flare loop.

Taken together, these results indicate that while electron beams can readily produce strong continuum enhancements and dynamic chromospheric responses, additional factors, such as deeper particle penetration, stronger magnetic fields, or cooler pre-flare atmospheric stratifications, may be required to reproduce the most extreme white-light and helioseismic signatures. Because the imposed beams are injected vertically rather than propagated explicitly along magnetic loops, the resulting energy deposition reflects the magnetic structuring of the atmosphere through its influence on column mass, but does not capture true field-aligned particle transport. The need for deeper penetration naturally directs attention toward high-energy proton beams \citep{Sadykov2024RADYN,2025ApJ...988...74G} and, perhaps, mixed electron-proton precipitation, which are capable of reaching larger column masses and potentially delivering the necessary photospheric heating. Similarly, the strong dependence of Fe\,I line response on local photospheric temperature highlights the importance of self-consistent pre-flare atmospheric conditions, particularly the cooler environments of sunspot umbrae where line reversal is observed.

\section{Conclusions}

The three-dimensional radiative MHD simulations presented here demonstrate that electron beams with parameters characteristic of strong flare electron-beam heating (energy flux density of $1\times10^{12}$ erg\,s$^{-1}$\,cm$^{-2}$, a spectral index of $\delta=3$, and $E_C$ between 10 and 25\,keV) produce substantial upper-chromospheric heating, strong continuum enhancement, and a complex hydrodynamic response (Figures \ref{B1_LogT}--\ref{z_2.25Mm}). The impulsive energy deposition generates multiple upward-propagating shock fronts, lateral expansions, and short-lived chromospheric condensations and bubbles, reflecting the inherently three-dimensional nature of flare energy transport. The simulated continuum brightens by a maximum factor of $\approx$ 2.2--2.5 and exhibits fine-scale structuring of the continuum emission (Figure \ref{NormIc}), indicating that electron beams can produce continuum enhancements comparable in magnitude to those observed in strong white-light flares and capture key aspects of their localized nature. However, the Fe\,I 6173\,\AA~line core does not transition into full emission, as observed in some strong flares \citep{Kosovichev2023, 2025ApJ...988...74G}, and remains in absorption for all tested beam parameters. The present models assume weak quiet-Sun magnetic fields and vertically imposed electron beams rather than field-aligned transport from reconnection sites. These simplifications may influence both the spatial distribution of heating and the achievable penetration depth and should be considered when comparing the results with strongly magnetized flare footpoints. In strongly magnetized flare loops, field-aligned transport and magnetic convergence may significantly modify the column-mass distribution and resulting energy deposition depths.

Comparison with one-dimensional F-CHROMA RADYN models highlights key differences that arise from multidimensional atmospheric structure. Despite identical beam parameters and substantial differences in pre-flare thermal stratification between the two modeling frameworks, the StellarBox atmospheres exhibit more rapid temperature rises, higher peak coronal temperatures, and faster post-impulsive cooling than their 1D counterparts (Figures \ref{Stel_vs_RAD_T} and \ref{Stel_vs_RAD_rho}). This behavior reflects the combined effects of lateral expansion and multidimensional redistribution of energy, rather than primarily reflecting differences in radiative loss prescriptions. While 1D models retain hot coronal plasma long after heating ceases, the three-dimensional simulations rapidly relax toward pre-flare conditions, suggesting that energy confinement may be overestimated in simplified geometries. These distinctions demonstrate the need for multidimensional modeling when interpreting flare energetics, particularly in the context of line formation and continuum response.

The synthetic Fe\,I 6173\,\AA~profiles further clarify the relationship between heating depth, atmospheric structure, and observable flare signatures. Line-core brightening is strongest above cooler intergranular lanes, where lower continuum background levels enhance the contrast of the partially filled-in line core. This sensitivity to local thermal structure mirrors HMI observations, which show Fe\,I emission kernels preferentially forming in umbral and umbra-penumbra regions. Because the present simulations sample quiet-Sun granulation rather than strong-field sunspot atmospheres, they do not capture the potentially greater susceptibility to line-core reversal expected from the cooler pre-flare stratifications of active-region atmospheres. The spatially fragmented continuum emission in our simulations may explain the fine structure of flare ribbons \citep{Sharykin2014,French2025}. The absence of full line-core reversal in the simulations suggests that significantly deeper and more concentrated heating, and/or different pre-flare atmospheric conditions, may be required to reproduce the observed emission. 

Taken together, these results suggest that additional energy-transport agents or atmospheric conditions, such as high-energy protons, mixed particle populations, or strong-field sunspot atmospheres, may play an important role in producing the deepest heating and most extreme flare signatures. However, a full assessment of electron-beam efficacy requires extending the present models to include umbral magnetic fields and cooler pre-flare stratifications. Future work will incorporate proton and mixed-species beams into the StellarBox framework, along with more realistic umbral and penumbral initial conditions, improved treatment of deep photospheric radiative transfer, and the inclusion of radiative-acoustic coupling needed to assess sunquake excitation.

\section{Appendix}
The electron-beam heating term implemented in StellarBox follows the analytic solution for collisional energy loss in a cold plasma \citep{1972SvA....16..273S, Livshits1981}.
At each horizontal location $(x,y)$ the vertical hydrogen column density $\xi(z)$ above height $z$ is
\begin{equation}
  d\xi = -n\,dz
  \label{eq:column}
\end{equation}
where $n(z)$ is the number density of neutral and ionized hydrogen atoms.

In the numerical implementation, the hydrogen column density in Equation~\ref{eq:column}
is evaluated discretely from the top of the domain downward. For each vertical cell $k$ with cell thickness $\Delta z_{k}$ and hydrogen number densities
$n_{k}=n(i,j,k)$ and $n_{k+1}=n(i,j,k+1)$ at adjacent cell centers, the incremental contribution to the cumulative column density is computed as:
\begin{equation}
  \xi_{k} = \xi_{k+1}
            + \frac{1}{2}\,
              \left(n_{k} + n_{k+1}\right)
              \,\Delta z_{k}.
  \label{eq:xidiscrete}
\end{equation}  
This trapezoidal summation accumulates
$\xi$ from the upper boundary
($\xi=0$ at $z_{\mathrm{top}}$) downward through the computational grid.

Next, we define characteristic column depth $\xi_0$ and ionization loss constant $a_e$ for electrons with energy $E$:
\begin{equation}
  \xi_{0} = \frac{E^{2}}{2a_{\mathrm{e}}}.
  \label{eq:xi0}
\end{equation}
\begin{equation}
  a_{\mathrm{e}}
   = 3.32\times10^{-37}
     \left[
       \ln\!\left(\frac{E}{m_ec^{2}}\right)
       -\tfrac{1}{2}\ln n
       +38.7
     \right],
  \label{eq:ae}
\end{equation}

The function of heating is:  
\begin{equation}
  P_{1}(\xi)
   = \frac{a_{\mathrm{e}}}{2}\,
     F_{0}\,E\,
     (2a_{\mathrm{e}}\xi)^{-3/2}\,
     \varphi(\xi),
  \label{eq:p1}
\end{equation}
where $F_{0}$ is the total beam energy flux density at the top of the atmosphere, and $\varphi$ is:

\begin{equation}
    \varphi (\xi) =
        \begin{cases}
        \frac{\pi}{2}, & \xi \ge \xi_{0},\\
        \frac{\pi}{2} - \arctan\left(\sqrt{\frac{\xi_{0}}{\xi}-1}\,\right) - \frac{\xi}{\xi_{0}} \sqrt{\frac{\xi_{0}}{\xi}-1}, & \xi < \xi_{0}.
        \end{cases}
    \label{eq:phi}
\end{equation}

The volumetric heating rate added to the MHD energy equation is
\begin{equation}
  Q_{\mathrm{beam}}(x,y,z,t)
   = P_{1}(\xi)\,n(z)\,
     \exp\!\left[-\frac{x^{2}+y^{2}}
                          {2\sigma_{\mathrm{beam}}^{2}}\right]
     f(t),
  \label{eq:qbeam}
\end{equation}
with a Gaussian lateral width $\sigma_{\mathrm{beam}}$ and a triangular temporal profile
\begin{equation}
  f(t)=
  \begin{cases}
    \dfrac{t-t_{0}}{0.5\,\Delta t},
      & 0\le t-t_{0}\le 0.5\,\Delta t,\\[6pt]
    \dfrac{\Delta t-(t-t_{0})}{0.5\,\Delta t},
      & 0.5\,\Delta t< t-t_{0}\le \Delta t,\\[6pt]
    0, & t-t_{0}>\Delta t,
  \end{cases}
  \label{eq:ftime}
\end{equation}
where $t_{0}$ is the beam onset time and $\Delta t$ its total duration.

\section*{Acknowledgment}
This work is partially supported by the NSF grant 1916509, and NASA Programs: the Science DRIVE (Diversify, Realize, Integrate, Venture, Educate) Center Program (COFFIES Project ``Consequences of Fields and Flows in the Interior and Exterior of the Sun''; 80NSSC22M0162), Heliophysics Guest Investigators (23-HGIO23\_2-0077), and Heliophysics Internal Scientist Funding (24-HISFM24\_2-0001 and 24-HISFM24\_2-0032). Resources supporting this work were provided by the NASA High-End Computing (HEC) Program through the NASA Advanced Supercomputing (NAS) Division at Ames Research Center. The authors acknowledge the use of the Overleaf Writeful and TeXGPT models for sentence structure and prose suggestions.


\begin{thebibliography}{}
\expandafter\ifx\csname natexlab\endcsname\relax\def\natexlab#1{#1}\fi
\providecommand{\url}[1]{\href{#1}{#1}}
\providecommand{\dodoi}[1]{doi:~\href{http://doi.org/#1}{\nolinkurl{#1}}}
\providecommand{\doeprint}[1]{\href{http://ascl.net/#1}{\nolinkurl{http://ascl.net/#1}}}
\providecommand{\doarXiv}[1]{\href{https://arxiv.org/abs/#1}{\nolinkurl{https://arxiv.org/abs/#1}}}

\bibitem[{{Allred} {et~al.}(2005){Allred}, {Hawley}, {Abbett}, \& {Carlsson}}]{Allred2005}
{Allred}, J.~C., {Hawley}, S.~L., {Abbett}, W.~P., \& {Carlsson}, M. 2005, \apj, 630, 573, \dodoi{10.1086/431751}

\bibitem[{{Allred} {et~al.}(2015){Allred}, {Kowalski}, \& {Carlsson}}]{2015ApJ...809..104A}
{Allred}, J.~C., {Kowalski}, A.~F., \& {Carlsson}, M. 2015, \apj, 809, 104, \dodoi{10.1088/0004-637X/809/1/104}

\bibitem[{{Brown}(1971)}]{1971SoPh...18..489B}
{Brown}, J.~C. 1971, \solphys, 18, 489, \dodoi{10.1007/BF00149070}

\bibitem[{{Carlsson} {et~al.}(2023){Carlsson}, {Fletcher}, {Allred}, {Heinzel}, {Ka{\v{s}}parov{\'a}}, {Kowalski}, {Mathioudakis}, {Reid}, \& {Sim{\~o}es}}]{Carlsson2023}
{Carlsson}, M., {Fletcher}, L., {Allred}, J., {et~al.} 2023, \aap, 673, A150, \dodoi{10.1051/0004-6361/202346087}

\bibitem[{{French} {et~al.}(2025){French}, {Kazachenko}, {Berghmans}, {D'Huys}, {Dominique}, {Patel}, {Talpeanu}, {Tamburri}, \& {Yadav}}]{French2025}
{French}, R.~J., {Kazachenko}, M.~D., {Berghmans}, D., {et~al.} 2025, arXiv e-prints, arXiv:2512.00710, \dodoi{10.48550/arXiv.2512.00710}

\bibitem[{{Frogner} \& {Gudiksen}(2024)}]{2024A&A...683A.195F}
{Frogner}, L., \& {Gudiksen}, B.~V. 2024, \aap, 683, A195, \dodoi{10.1051/0004-6361/202348457}

\bibitem[{{Frogner} {et~al.}(2020){Frogner}, {Gudiksen}, \& {Bakke}}]{2020A&A...643A..27F}
{Frogner}, L., {Gudiksen}, B.~V., \& {Bakke}, H. 2020, \aap, 643, A27, \dodoi{10.1051/0004-6361/202038529}

\bibitem[{{Granovsky} {et~al.}(2025){Granovsky}, {Kosovichev}, {Sadykov}, {Kerr}, \& {Allred}}]{2025ApJ...988...74G}
{Granovsky}, S., {Kosovichev}, A.~G., {Sadykov}, V.~M., {Kerr}, G.~S., \& {Allred}, J.~C. 2025, \apj, 988, 74, \dodoi{10.3847/1538-4357/addd1e}

\bibitem[{{Hudson}(1972)}]{1972SoPh...24..414H}
{Hudson}, H.~S. 1972, \solphys, 24, 414, \dodoi{10.1007/BF00153384}

\bibitem[{{Imada} {et~al.}(2013){Imada}, {Aoki}, {Hara}, {Watanabe}, {Harra}, \& {Shimizu}}]{2013ApJ...776L..11I}
{Imada}, S., {Aoki}, K., {Hara}, H., {et~al.} 2013, \apjl, 776, L11, \dodoi{10.1088/2041-8205/776/1/L11}

\bibitem[{{Kosovichev} {et~al.}(2023){Kosovichev}, {Sadykov}, \& {Stefan}}]{Kosovichev2023}
{Kosovichev}, A.~G., {Sadykov}, V.~M., \& {Stefan}, J.~T. 2023, \apj, 958, 160, \dodoi{10.3847/1538-4357/acf9eb}

\bibitem[{{Kostiuk} \& {Pikelner}(1975)}]{1975SvA....18..590K}
{Kostiuk}, N.~D., \& {Pikelner}, S.~B. 1975, \sovast, 18, 590

\bibitem[{{Livshits} {et~al.}(1981){Livshits}, {Badalian}, {Kosovichev}, \& {Katsova}}]{Livshits1981}
{Livshits}, M.~A., {Badalian}, O.~G., {Kosovichev}, A.~G., \& {Katsova}, M.~M. 1981, \solphys, 73, 269, \dodoi{10.1007/BF00151682}

\bibitem[{{Machado} {et~al.}(1989){Machado}, {Emslie}, \& {Avrett}}]{1989SoPh..124..303M}
{Machado}, M.~E., {Emslie}, A.~G., \& {Avrett}, E.~H. 1989, \solphys, 124, 303, \dodoi{10.1007/BF00156272}

\bibitem[{{Monson} {et~al.}(2021){Monson}, {Mathioudakis}, {Reid}, {Milligan}, \& {Kuridze}}]{Monson2021}
{Monson}, A.~J., {Mathioudakis}, M., {Reid}, A., {Milligan}, R., \& {Kuridze}, D. 2021, \apj, 915, 16, \dodoi{10.3847/1538-4357/abfda8}

\bibitem[{{Pereira} \& {Uitenbroek}(2015)}]{pereira2015RH15D}
{Pereira}, T. M.~D., \& {Uitenbroek}, H. 2015, \aap, 574, A3, \dodoi{10.1051/0004-6361/201424785}

\bibitem[{{Reid} {et~al.}(2020){Reid}, {Zhigulin}, {Carlsson}, \& {Mathioudakis}}]{2020ApJ...894L..21R}
{Reid}, A., {Zhigulin}, B., {Carlsson}, M., \& {Mathioudakis}, M. 2020, \apjl, 894, L21, \dodoi{10.3847/2041-8213/ab8d1e}

\bibitem[{{Rybicki} \& {Hummer}(1991)}]{rybicki1991RHI}
{Rybicki}, G.~B., \& {Hummer}, D.~G. 1991, \aap, 245, 171

\bibitem[{{Rybicki} \& {Hummer}(1992)}]{rybicki1992RHII}
---. 1992, \aap, 262, 209

\bibitem[{{Sadykov} {et~al.}(2020){Sadykov}, {Kosovichev}, {Kitiashvili}, \& {Kerr}}]{Sadykov2020}
{Sadykov}, V.~M., {Kosovichev}, A.~G., {Kitiashvili}, I.~N., \& {Kerr}, G.~S. 2020, \apj, 893, 24, \dodoi{10.3847/1538-4357/ab7b6a}

\bibitem[{{Sadykov} {et~al.}(2024){Sadykov}, {Stefan}, {Kosovichev}, {Stejko}, {Kowalski}, {Allred}, \& {Kerr}}]{Sadykov2024RADYN}
{Sadykov}, V.~M., {Stefan}, J.~T., {Kosovichev}, A.~G., {et~al.} 2024, \apj, 960, 80, \dodoi{10.3847/1538-4357/ad0cf3}

\bibitem[{{Scherrer} {et~al.}(2012){Scherrer}, {Schou}, {Bush}, {Kosovichev}, {Bogart}, {Hoeksema}, {Liu}, {Duvall}, {Zhao}, {Title}, {Schrijver}, {Tarbell}, \& {Tomczyk}}]{Scherrer2012}
{Scherrer}, P.~H., {Schou}, J., {Bush}, R.~I., {et~al.} 2012, \solphys, 275, 207, \dodoi{10.1007/s11207-011-9834-2}

\bibitem[{{Sharykin} \& {Kosovichev}(2014)}]{Sharykin2014}
{Sharykin}, I.~N., \& {Kosovichev}, A.~G. 2014, \apjl, 788, L18, \dodoi{10.1088/2041-8205/788/1/L18}

\bibitem[{{Syrovatskii} \& {Shmeleva}(1972)}]{1972SvA....16..273S}
{Syrovatskii}, S.~I., \& {Shmeleva}, O.~P. 1972, \sovast, 16, 273

\bibitem[{{Uitenbroek}(2001)}]{uitenbroek2001RHPRD}
{Uitenbroek}, H. 2001, \apj, 557, 389, \dodoi{10.1086/321659}

\bibitem[{{Wray} {et~al.}(2015){Wray}, {Bensassi}, {Kitiashvili}, {Mansour}, \& {Kosovichev}}]{2015arXiv150707999W}
{Wray}, A.~A., {Bensassi}, K., {Kitiashvili}, I.~N., {Mansour}, N.~N., \& {Kosovichev}, A.~G. 2015, arXiv e-prints, arXiv:1507.07999, \dodoi{10.48550/arXiv.1507.07999}

\end{thebibliography}

\end{document}